# Optimization Approach to Parametric Tuning of Power System Stabilizer Based on Trajectory Sensitivity Analysis

Yuan Zhang, *Graduate Student Member, IEEE*

***Abstract*—This paper proposed an transient-based optimal parametric tuning method for power system stabilizer (PSS) based on trajectory sensitivity (TS) analysis of hybrid system, such as hybrid power system (HPS). The main objective is to explore a systematic optimization approach of PSS under large disturbance of HPS, where its nonlinear features cannot be ignored, which, however, the traditional eigenvalue-based small signal optimizations do neglect the higher order terms of Taylor series of the system state equations. In contrast to previous work, the proposed TS optimal method focuses on the gradient information of objective function with respect to decision variables by means of the trajectory sensitivity of HPS to the PSS parameters, and optimizes the PSS parameters in terms of the conjugate gradient method. Since the objective function considers the transient features under large disturbances of HPS, the proposed method could be used to inhibit the spontaneous oscillation caused by small disturbance and to effectively damp the system oscillation caused by large disturbance. Firstly, the application background and traditional parametric tuning optimization methods of PSS are introduced. Then, the systematic mathematical models and transient trajectory simulation are presented by introducing switching/reset events in terms of triggering hypersurfaces so as to formulate the optimization problem using TS analysis. Finally, a case study of IEEE three-machine-nine-bus standard test system is discussed in detail to exemplify the practicality and effectiveness of the proposed optimal method.**

## I. Introduction

THE GRID INTERACONNECTION of large power system can economically increase the operation efficiency, which however, will also lead to the low-frequency self-oscillation of the power system that inhibits its long-term stability [1]. The power system stabilizer (PSS) is introduced as a feedback controller to decrease such oscillation, and is very helpful to increase the reliability of large power system. Due to the improperly designed values of some practically relevant parameters, distortion behaviors can frequently occur, which will lead to the rapid rise of device stress and the drastic degeneration of power system performance [2]. Hence, the optimal parametric tuning issue in terms of some major parameters of the PSS feedback controller is very crucial to the power system operation, and has become an important goal and subject of much on-going research [3].

Up to now, several typical approaches such as phase compensation method, based method, linear optimization method and neural network PSS optimization method have been proposed to explore the optimal parametric tuning policy for PSS [4]. Although these previous results have greatly improved the optimization methods of PSS, they merely focused on small signal behaviors under small disturbance of power system [5]. However, a large power system is essentially a hard (nonlinear and nonsmooth) dynamic system, where the oscillation under large disturbance is totally different from the small signal-based self-excited oscillation. By using a normal form technique, C. M. Lin *et al.*, have illustrated that the mechanism of power system undergoing large disturbance, which there exist both the fundamental oscillation patterns and high-order oscillation ones [6]. Since the traditional eigenvalue-based small signal optimization methods didn't consider the higher order terms of Taylor series of the system state equation, its results in the fact that some useful information essential to design-oriented optimization is of great difficulty to obtain. Thus, these small signal-based optimization methods might fail to obtain a globally optimal parameter set of PSS, namely, they are not able to significantly damp oscillations caused by large disturbance in the power system. Moreover, due to the complexity of power system dynamics, values that are optimal for one disturbance scenario may be suboptimal for others. Thus, a systematic and optimization-based PSS parameter tuning approach should be address for each case.

Unfortunately, even under the aforementioned large disturbance case, the parametric PSS optimization associated with system modeling could be rather difficult due to discontinuous change of power system structures, which generally possess nonlinear and nonsmooth transient features [7]. For example, for a three short circuit at a key bus, the topology of original power network together with its nodes voltage amplitudes and angles will be jumped to another pattern. As far as a large power system of great complexity and high dimension are concerned, the system optimization analysis will become more complicate, which doesn't merely possess continuous-time dynamics. In fact, such a large physical system exhibit dynamic behavior which is governed by a mix of continuous-time (possibly constrained) dynamics, discrete-time and discrete-event dynamics, switching action, and jump phenomena, that has become known generically a hybrid systems [3], [8]-[10]. The optimization problem of such hybrid power system should be governed by a set of nonlinear dynamics with many components, including generators, loads, flexible AC transmission system

This work was supported by the Dean's Fellowship Program of Boston University. Y. Zhang is with the Division of System Engineering, College of Engineering, Boston University, 15 Saint Mary's Street, Brookline, MA 02446 USA. E-mail: yzboston@bu.edu.

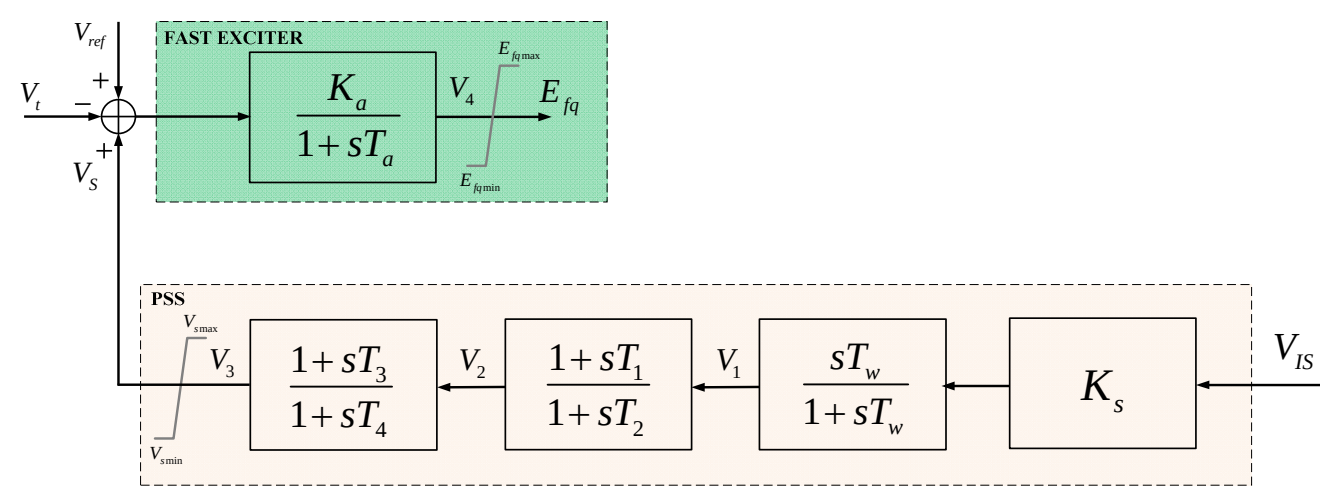

Fig. 1. Feedback control block diagram of PSS with a thyristor-type fast exciter.

(FACTS) devices, and their associated control devices [3]. Therefore, power system behaviors can be quite complicated, which lead to the fact that a better parametric optimization of PSS and other equipments should be reliant on a thorough understanding of those behaviors.

As indicated in [3]-[12], the analysis of most hybrid system is generally reliant on time-domain simulation. Compared with few analytical method such as Lyapunov-type theory for specific hybrid system applications [13-15], the major advantage of simulation is that it is applicable for arbitrarily complicated models, which however, can merely provide information about a single scenario and could involve large computational costs for large power systems. To compensate such drawback, trajectory sensitivity (TS) analysis for general hybrid system has been proposed recently to offer an effective and insightful method on the system transient evolution [3], [16]-[17]. Different from analyzing in the neighborhood of an equilibrium point or a periodic cycle, TS approach focus on linearization around a nonlinear, and possibly non-smooth, trajectory, which could be transient or steady [3]. Hence, the influence of parameter variations on large disturbance behavior can be estimated (to first order) from their TSs. In the above sense, the actual concept of PSS parametric tuning should be based on certain transient-based optimization approach to evaluate the TS of power system with respect to these parameters under the constraints of the corresponding hybrid system model.

In this paper, an optimization-based model is proposed to extend the aforementioned TS method to the parametric tuning of PSS under a large disturbance of power system. The proposed method begins with a hybrid system model in the form of modified differential-algebraic-discrete (DAD) equation that can be regarded as a generic model of hybrid power system (HPS) under large disturbance. Unlike the previous studies, it focuses on the gradient information of a proposed objective function with respect to decision variables by means of the system trajectory sensitivity to these PSS parameters, and optimizes the PSS parameters in terms of the conjugate gradient method. For readability and effectiveness of exposition, we illustrate the method and its salient features using IEEE three-machine-nine-bus standard test system as an example. The analysis is of help to us in doping out the influence of PSS parameters on the dynamic behaviors of power system undergoing large disturbance, and more significantly, some optimal decision variables obtained here can facilitate design of PSS feedback controller for effectively damp the system oscillation.

The paper is organized as follows. Section II gives a brief explanation of PSS and a detailed mathematical model of HPS. Section III formulates the optimization problem for PSS parametric tuning by using TS, where its constraints are built in the compact model of HPS that suffers discontinuity. In Section IV, based on the proposed TS-based optimal parametric tuning approach, optimal PSS parameters are analyzed as an example to exemplify the proposed method by using IEEE standard test system, where some interesting results and analysis are illustrated. Finally, some remarkable conclusions are arrived at in Section V.

## II. Modeling of PSS and HPS

In this section, the mathematical models of PSS and HPS are presented, respectively. The systematic discontinuous behavior of HPS undergoing discrete-events is analyzed in details, by considering both switching event and reset event. It indicates that the TS information of HPS under large disturbance could be helpful to optimize parametric tuning of PSS.

### A. Configurations of PSS with Exciter

Generator excitation systems (with PSS) play a fundamental role in power system control. Fig. 1 Provides a typical feedback control block diagram of PSS with a thyristor-type fast exciter, where $K_a$ and $T_a$ are the gain and time constant of the exciter, respectively; $K_s$ is the gain of PSS; $T_w$ is the time constant of DC blocking component for PSS; $T_1$, $T_2$, $T_3$, $T_4$ represent as the time constant of leading and lagging part for PSS, respectively; $E_{fd}$, $V_S$ are denoted as the output of exciter and PSS, respectively; $V_{IS}$ is denoted as the input of PSS; $V_t$ represents the same input of exciter and PSS; $V_{ref}$ is the reference voltage of exciter; the subscript *max* and *min* are denoted as the upper and lower bound for their corresponding variables [18]. Here, all variables and parameters are per-unit values. Subsequently, we can derive the differential equations for the above PSS block.

$$\frac{d[T_w(K_s V_{IS} - V_1)]}{dt} = V_1 \tag{1}$$

$$\frac{d(T_2 V_2 - T_1 V_1)}{dt} = V_1 - V_2 \tag{2}$$

$$\frac{d(T_4 V_2 - T_3 V_1)}{dt} = V_2 - V_3 \tag{3}$$

$$\frac{d(T_a V_4)}{dt} = K_a(V_{ref} + V_s - V_t) - V_4 \tag{4}$$

Fig. 1 shows that the two leading and lagging part in PSS can be regarded as identical. Thus, in our later optimal analysis, we consider both parameters are the same, namely, $T_1 = T_3$, $T_2 = T_4$. Since $K_w$ is usually determined by experience as indicated in [18], thus the parametric tuning of PSS is define in the parameter set $\boldsymbol{\lambda} = (K_s, T_1, T_2, T_3, T_4)$. For the input of PSS, we usually select the rotational speed $\omega_i$ of the *i*-th generator that needs to be controlled since $\omega_i$ can be easily measured. It has shown that the system can achieve a quite good dynamic response by using $\omega_i$ as the input signal. In this paper, we implement such state variable as our input variable. We will shortly see that the TS information of power system under large disturbance can be obtained from $\omega_i$.

## B. Mathematical Model of HPS

As illustrated previously, a power system should be consider as a hybrid system, which involves a mix of continuous-time dynamics, discrete-time and discrete-event dynamics, switching action, and jump phenomena. It can be generally described by a set of parameter-dependent DAD models, adapted to incorporate impulsive (state reset) action and switching of the algebraic equations [19], [20]. A generic HPS containing multi-generators can be usually decomposed into four major parts, i.e., generators, loads, power networks and the aforementioned PSS components as exciter part. For readability and effectiveness of exposition, we give two definitions of *switching event* and *reset event* which play a vital role in the discrete-event cases that could emerge in the HPS, which was ignored in [3].

*Definition 2.1*: (*Switching Event*) Switching event $SE^{(i)}$ is defined as any event that can directly trigger the change of algebraic states $\boldsymbol{y} \in \mathbb{R}^m$ at the *i*-th period, which can then form a switching event set $A_{SE}$, with its index set denoted as $I_{A_{SE}}$.

*Remark 2.1*: For the aforementioned HPS, a switching event $SE^{(i)}$ can be the open circuit and short circuit cases of the power network as shown in Fig. 3. Under both switching events, some algebraic states $\boldsymbol{y}$, e.g., load bus voltage magnitudes and angles, injection current from generator to power network, generator electromagnetic power, are forced to be changed. To satisfy the physical constraints of Kirchhoff's laws, other algebraic variables must undergo a step change.

*Definition 2.2*: (*Reset Event*) Reset event $RE^{(j)}$ is defined as any event that can directly trigger the change of discrete states $\boldsymbol{z} \in \mathbb{R}^l$ at the *j*-th period, which can then form a reset event set $A_{RE}$ with it index set denoted as $I_{A_{RE}}$.

*Remark 2.2*: Likewise, for the HPS, the discrete states $\boldsymbol{z}$ can be regarded as transformer tap positions and/or protection relay logic states as shown in Fig. 4 with red highlights. In the case of modifying transformer tap positions, the algebraic states $\boldsymbol{y}$ may again step to ensure the physical constraints of the power network.

Note that the above two discrete-events can both lead to the change algebraic states $\boldsymbol{y}$ and discrete states $\boldsymbol{z}$. However, it should be emphasized the continuous dynamic states $\boldsymbol{x}_c \in \mathbb{R}^n$ of any real HPS cannot undergo step change due to the continuity of the solution with respect to initial conditions for differential equations if the Lipschitz continuity satisfies.

Therefore, a compact form of parameter-dependent DAD model for the HPS can be written as follows.

$$\dot{\boldsymbol{x}} = \boldsymbol{f}(\boldsymbol{x}, \boldsymbol{y}) \tag{5}$$

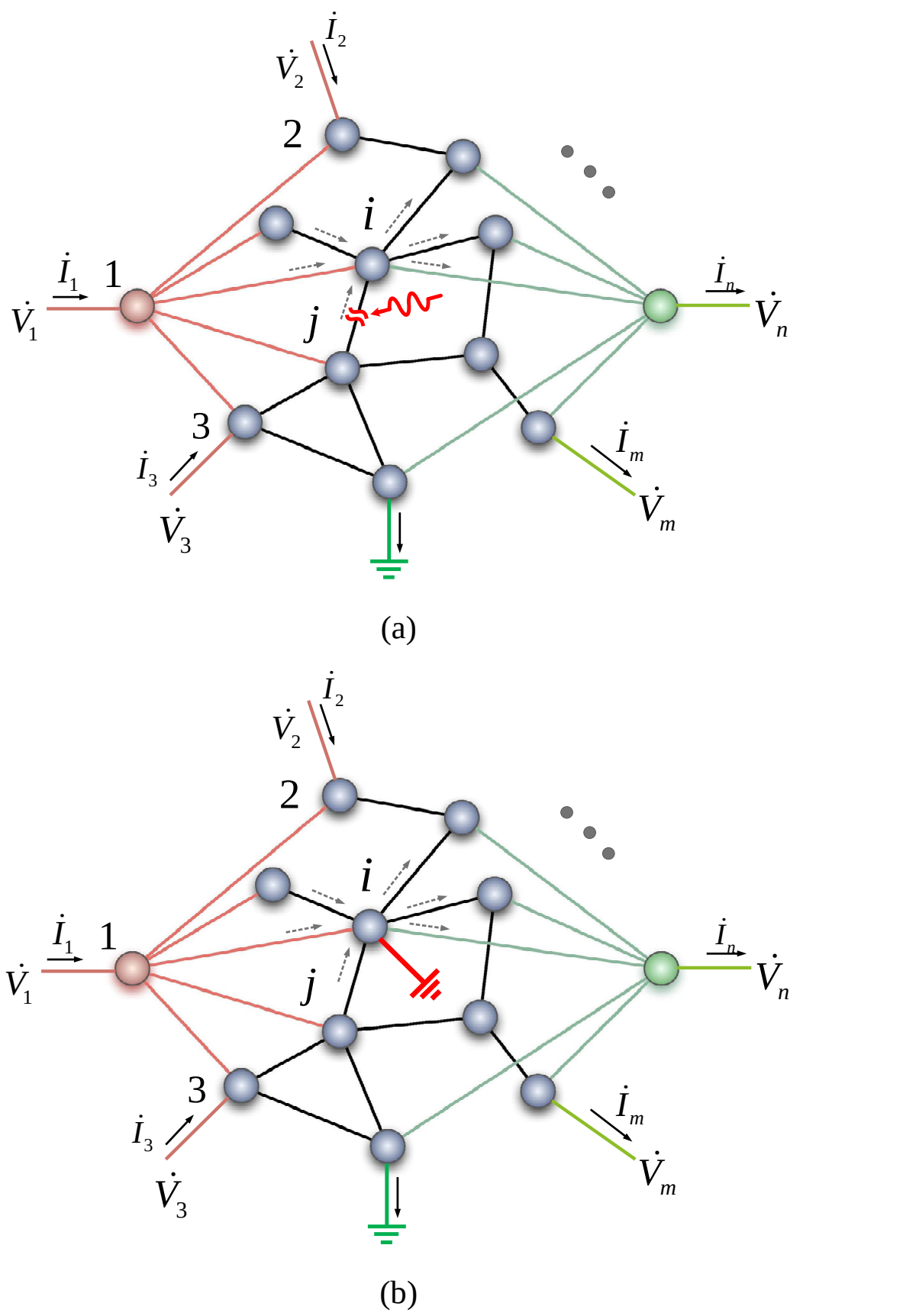

Fig. 2. Configurations of power network incorporated with switching events. (a) Open circuit between major node $i$ and $j$; (b) Short circuit at major node $j$.

$$\boldsymbol{0} = \boldsymbol{g}^{(0)}(\boldsymbol{x}, \boldsymbol{y}) \tag{6}$$

$$\boldsymbol{0} = \begin{cases} \boldsymbol{g}^{(i-)}(\boldsymbol{x}, \boldsymbol{y}), & SE^{(i)} \notin A_{SE}; \\ \boldsymbol{g}^{(i+)}(\boldsymbol{x}, \boldsymbol{y}), & SE^{(i)} \in A_{SE}; \end{cases} \quad \forall i \in I_{A_{SE}} \tag{7}$$

$$\boldsymbol{z}^+ = \boldsymbol{h}^{(j)}(\boldsymbol{x}^-, \boldsymbol{y}^-), \quad RE^{(j)} \in A_{RE}; \quad \forall j \in I_{A_{RE}} \tag{8}$$

$$\dot{\boldsymbol{z}} = \boldsymbol{0}, \quad RE^{(j)} \notin A_{RE}; \quad \forall j \in I_{A_{RE}} \tag{9}$$

where $\boldsymbol{x} = [\boldsymbol{x}_c, \boldsymbol{z}, \boldsymbol{\lambda}] \in \mathbb{R}^{n+m+p}$, $\boldsymbol{f} : \mathbb{R}^{n+l+p+m} \to \mathbb{R}^n$, $\boldsymbol{g} : \mathbb{R}^{n+l+p+m} \to \mathbb{R}^m$, $\boldsymbol{h}^{(j)} : \mathbb{R}^{n+l+p+m} \to \mathbb{R}^l$ and $\boldsymbol{\lambda} \in \mathbb{R}^p$ represents systematic parameters such as generator reactance, controller gains and switching times, including the aforementioned PSS control parameters. $\boldsymbol{x}^-$, $\boldsymbol{y}^-$ refer to the values of $\boldsymbol{x}$, $\boldsymbol{y}$ just prior to the reset condition, while $\boldsymbol{z}^+$ denotes the value of $\boldsymbol{z}$ just after the reset event. Also, $i-$, $i+$ denotes the previous and later switching events, respectively. Here, by incorporating parameters $\boldsymbol{\lambda}$ into the state $\boldsymbol{x}$ x allows a convenient development of trajectory sensitivities. For the convenience of the following discussions, we denote the above system as $\boldsymbol{\Gamma}_1$, and will follow these notation in the following discussions.

The above model is similar to a model proposed in [3], but with a more general expression of the two discrete-events. This HPS model can captures all the important aspects of its behavior, namely, the interaction between continuous, algebraic and discrete states as they are driven by time and/or event. Moreover, as mentioned previously, the main difficulty of analyzing the HPS lies in the discontinuity and discrete states which caused at the *jump time* (or denoted as *junction time*) between different DA systems. Unfortunately, we are now just face to this obstacle since the original idea of implementing PSS is to inhibit certain side-effects when the HPS encounters the aforementioned discrete-events. Thus, an optimization-based parametric tuning of PSS should be reliant on fully understanding of the system behavior at events and the intervals between various switching/reset events. Accordingly, we can map the HPS into the following simplified form by treating the events $SE^{(i)}$ and $RE^{(j)}$ as the case when the states trajectories encounter two different *triggering hypersurfaces* $H^{(i)}(\boldsymbol{x},\boldsymbol{y})$ and $S^{(j)}(\boldsymbol{x},\boldsymbol{y})$, respectively.

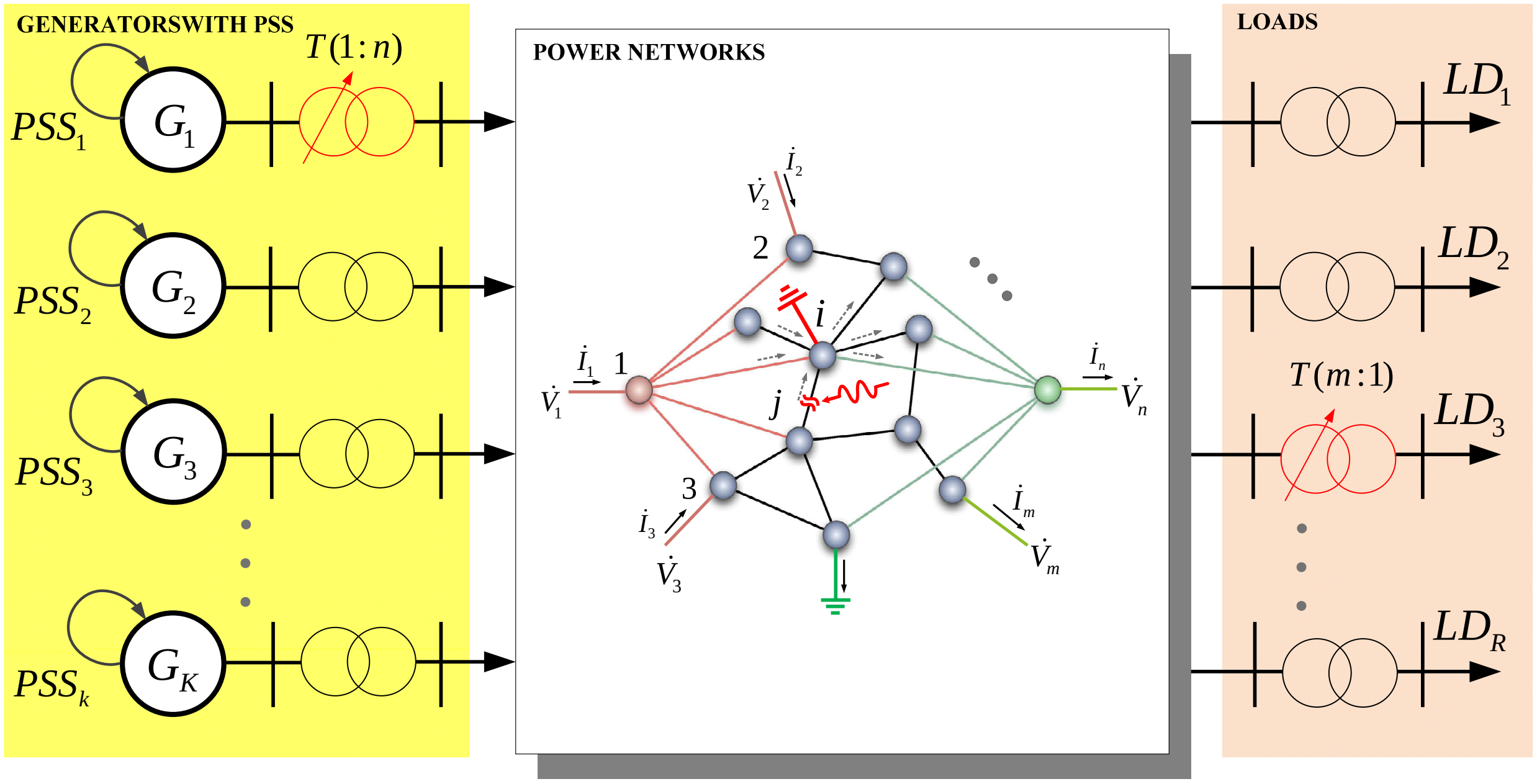

Fig. 3. HPS diagram incorporating with both switching events and reset events.

$$\dot{\boldsymbol{x}} = \boldsymbol{f}(\boldsymbol{x},\boldsymbol{y}) \tag{10}$$

$$\boldsymbol{0} = \boldsymbol{g}^{(0)}(\boldsymbol{x},\boldsymbol{y}) \tag{11}$$

$$\boldsymbol{0} = \begin{cases} \boldsymbol{g}^{(i-)}(\boldsymbol{x},\boldsymbol{y}), & H^{(i)}(\boldsymbol{x},\boldsymbol{y}) > 0; \\ \boldsymbol{g}^{(i+)}(\boldsymbol{x},\boldsymbol{y}), & H^{(i)}(\boldsymbol{x},\boldsymbol{y}) < 0; \end{cases} \quad \forall i \in I_{ASE} \tag{12}$$

$$\boldsymbol{z}^{+} = \boldsymbol{h}^{(j)}(\boldsymbol{x}^{-},\boldsymbol{y}^{-}), \quad S^{(j)}(\boldsymbol{x},\boldsymbol{y}) > 0; \quad \forall j \in I_{ARE} \tag{13}$$

$$\dot{\boldsymbol{z}} = \boldsymbol{0}, \quad S^{(j)}(\boldsymbol{x},\boldsymbol{y}) < 0; \quad \forall j \in I_{ARE} \tag{14}$$

It is worth mentioning that the switching $SE^{(i)}$ and reset event $RE^{(j)}$ are triggered by the conditions $H^{(i)}(\boldsymbol{x},\boldsymbol{y}) = 0$ and $S^{(j)}(\boldsymbol{x},\boldsymbol{y}) = 0$. In this paper, we will restrict ourselves in the situation that these events can be fully described by a condition using algebraic state $\boldsymbol{y}$ under a typical short-circuit case as shown in Section IV, and we will consider the HPS at two switching events at the junction epoch $t_{J1}$, $t_{J2}$, which represent the occurrence and disappearance of short-circuit cases. Thus, attention is focused on the simplified HPS model

$$\dot{\boldsymbol{x}} = \boldsymbol{f}(\boldsymbol{x},\boldsymbol{y}) \tag{15}$$

$$\boldsymbol{0} = \boldsymbol{g}^{(0)}(\boldsymbol{x},\boldsymbol{y}) \tag{16}$$

$$\boldsymbol{0} = \begin{cases} \boldsymbol{g}^{(i-)}(\boldsymbol{x},\boldsymbol{y}), & H^{(i)}(\boldsymbol{x},\boldsymbol{y}) > 0; \\ \boldsymbol{g}^{(i+)}(\boldsymbol{x},\boldsymbol{y}), & H^{(i)}(\boldsymbol{x},\boldsymbol{y}) < 0; \end{cases} \quad \forall i \in \{1,2\} \tag{17}$$

$$\dot{\boldsymbol{z}} = \boldsymbol{0}, \quad S^{(j)}(\boldsymbol{x},\boldsymbol{y}) < 0; \quad \forall j \in I_{ARE} \tag{18}$$

For simplicity, we denote the above HPS system (15)-(18) as $\boldsymbol{\Gamma}_2$. Generally, the flow of $\boldsymbol{x}$ and $\boldsymbol{y}$ can be used to describe the trajectory behavior of $\boldsymbol{\Gamma}_2$ as follows.

$$\boldsymbol{x}(t) = \boldsymbol{\psi}_{\boldsymbol{x}}(\boldsymbol{x}_o, t) \tag{19}$$

$$\boldsymbol{y}(t) = \boldsymbol{\psi}_{\boldsymbol{y}}(\boldsymbol{x}_o, t) \tag{20}$$

along with their initial conditions

$$\boldsymbol{x}(t_0) = \boldsymbol{\psi}_{\boldsymbol{x}}(\boldsymbol{x}_o, t_0) \equiv \boldsymbol{x}_o \tag{21}$$

$$\boldsymbol{0} = \boldsymbol{g}(\boldsymbol{x}_o, \boldsymbol{\psi}_{\boldsymbol{y}}(\boldsymbol{x}_o, t_0)) \equiv \boldsymbol{g}(\boldsymbol{x}_o, \boldsymbol{y}_o) \tag{22}$$

$$\dot{\boldsymbol{\lambda}} = \boldsymbol{0} \tag{23}$$

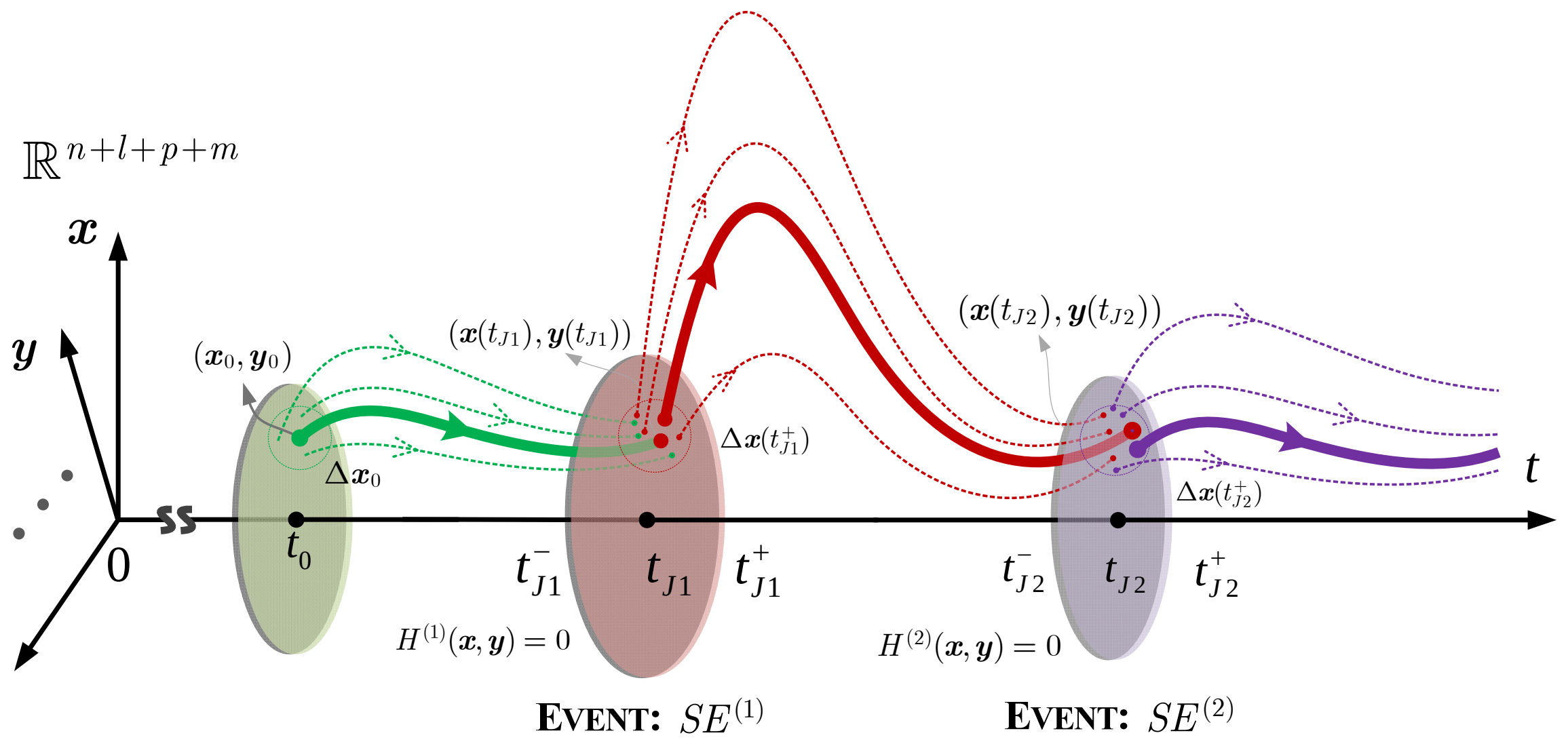

Fig. 4. Graphical view of jump conditions for HPS ( $\boldsymbol{\Gamma}_1$ ) undergoing two discrete events.

where $\boldsymbol{y}_o$ can be obtain from (22) based on the implicit function theorem as indicated in [3]. To ensure that parameter $\boldsymbol{\lambda}$ remain fixed at their initial values, the corresponding differential equations are defined as (23). Here, we only assume that $\boldsymbol{y}_o$ can be express as $\boldsymbol{x}_o$ and $t_0$ which is often satisfied in real HPS. Also, we assume that the HPS has already operated in steady condition before any event occurs. The initial conditions before the events can be obtained by using power flow calculation [7]. It is worth mentioning that $\boldsymbol{x}_o$ includes the initial conditions of the system parameters, including the PSS ones, where a graphical view of such jump condition of HPS has been depicted in Fig. 4. Since the main purpose of this paper elaborates in the systematic modeling and optimization analysis, we will not focus on concrete explanation of each component of equation (15)-(18). For conciseness, we refer the readers to [7] for a detailed illustration.

## III. Optimal Parametric Tuning of PSS based on TS

In Section II, a compact form of the HPS model has been developed. We will use it as constraints to formulate the optimization problem of PSS parametric tuning, the focus of this paper, which is also one application of optimization in the analysis of power system dynamics.

### A. Formulation of Optimal Parametric Tuning of PSS

As indicated in [18], for most optimization problems with constrains of the form of differential equations, a Bolza form in the classical calculus of variations on the conditions for an extremum, can be formulated using the following objective function

$$\min_{\boldsymbol{\lambda}} = J(\boldsymbol{x}, \boldsymbol{y}, t_f)$$

$$\text{s.t.} \quad J = \varphi(\boldsymbol{x}(t_f), \boldsymbol{y}(t_f), t_f) + \int_{t_0}^{t_f} \psi(\boldsymbol{x}(t), \boldsymbol{y}(t), t) dt \tag{24}$$

where $t_f$ is the final time a triggering event. The objective of PSS parametric tuning is to coordinate the function of PSS for multi-generators to achieve an optimal damping so as to force the HPS to recover to the post-disturbance stable operating point as quickly as possible. Generator rotational speed $w_i$ of the $i$-th generator can typically enable a good assessment of that recovery. An illustration has already been provided in Fig. 4, where a significant rotational speed deviation is an indicator of marginal system stability during the transient period following a large disturbance. Therefore, PSS parameters that minimize rotational speed deviation can effectively maximize system recovery. In this paper, we will use $w_i$ to formulate our objective function, and optimizes the PSS parameters in terms of the conjugate gradient method by obtaining its gradient information with respect to PSS parameters from TS analysis.

Hence, the objective function for optimizing PSS parameters can be formulated as

$$\begin{aligned}
&\min_{\boldsymbol{\lambda}} \quad J(\boldsymbol{x}, t_f) = \sum_{i=1}^{K} \int_{t_0}^{t_f} \omega_i^2 dt \\
\text{s.t.} \quad & \dot{\boldsymbol{x}} = \boldsymbol{f}(\boldsymbol{x}, \boldsymbol{y}) \\
& \boldsymbol{0} = \boldsymbol{g}^{(0)}(\boldsymbol{x}, \boldsymbol{y}) \\
& \boldsymbol{0} = \begin{cases} \boldsymbol{g}^{(i-)}(\boldsymbol{x}, \boldsymbol{y}), & H^{(i)}(\boldsymbol{x}, \boldsymbol{y}) > 0; \\ \boldsymbol{g}^{(i+)}(\boldsymbol{x}, \boldsymbol{y}), & H^{(i)}(\boldsymbol{x}, \boldsymbol{y}) < 0; \end{cases} \quad \forall i \in \{1, 2\} \\
& \dot{\boldsymbol{z}} = \boldsymbol{0}, \quad S^{(j)}(\boldsymbol{x}, \boldsymbol{y}) < 0; \quad \forall j \in I_{ARE} \\
& \boldsymbol{x}(t_0) = \boldsymbol{x}_o \\
& \boldsymbol{y}(t_0) = \boldsymbol{y}_o \\
& \underline{\boldsymbol{\lambda}}_i \leq \boldsymbol{\lambda}_i \leq \overline{\boldsymbol{\lambda}}_i, \quad \forall i \in \{1, 2, ..., K\}
\end{aligned} \tag{25}$$

where $\boldsymbol{\lambda}=\{\boldsymbol{\lambda}_1,\boldsymbol{\lambda}_2,\cdots,\boldsymbol{\lambda}_k\}$ , $\boldsymbol{\lambda}_i=\{K_{si},T_{1i},T_{2i}\}$ ; $K$ is the number of generators; $t_0$ and $t_f$ denote the beginning epoch of triggering event /large disturbance and ending epoch of simulation time, respectively, where $t_f>t_{J2}>t_{J1}>t_0$ . Note that $\boldsymbol{x}$ contains the decision variable $\boldsymbol{\lambda}$ in (25). For the objective function, it maps the trajectory of generator rotational speed $\omega_i$ into a scalar function $J(\boldsymbol{x},t_f)$, which has the form of a continuous-time nonlinear least-square. It has been indicated in [21] that if $J(\boldsymbol{x},t_f)$ is a smooth function of its arguments, then it is continuously differentiable in terms of $\boldsymbol{\lambda}$ and $t_f$. For the constraints shown in (25) , $\boldsymbol{\lambda}$ belongs to the convex set $[\underline{\boldsymbol{\lambda}}_i,\bar{\boldsymbol{\lambda}}_i]$, and other algebraic states in $\boldsymbol{x}$ undergo discontinuity over the two triggering events.

The minimization can therefore be solved using gradient-based methods, which require gradient information $\nabla_{\boldsymbol{\lambda}}J(\boldsymbol{x},t_f)$ with an explicit form as follows.

$$\nabla_{\boldsymbol{\lambda}}J(\boldsymbol{x},t_f)=\sum_{t_o}^{t_f}\sum_{i=1}^{n}2\omega_i(\nabla_{\boldsymbol{\lambda}}\omega_i)\Delta t \tag{26}$$

*B. TS Analysis for HPS*

For (26), the gradient information $\nabla_{\boldsymbol{\lambda}}J(\boldsymbol{x},t_f)$ can be obtained by analyzing another gradient (or denoted as *sensitivity* ) information of state variable $\omega_i$ with respect to $\boldsymbol{\lambda}$ , namely $\nabla_{\boldsymbol{\lambda}}\omega_i$ . A better formulation of this *sensitivity* information can be generally derived by taking the Talyor series expansion of (19), (20) as follows [3].

$$\Delta\boldsymbol{x}(t)=\nabla'_{\boldsymbol{x}_0}\boldsymbol{x}(t)\Delta\boldsymbol{x}_0 \tag{27}$$

$$\Delta\boldsymbol{y}(t)=\nabla'_{\boldsymbol{x}_0}\boldsymbol{y}(t)\Delta\boldsymbol{x}_0 \tag{28}$$

where $\nabla_{\boldsymbol{x}_0}\boldsymbol{x}$ , $\nabla_{\boldsymbol{x}_0}\boldsymbol{y}$ denote as the *TS* of trajectory $\boldsymbol{x}(t)$ and $\boldsymbol{y}(t)$, respectively. Recall that $\boldsymbol{x}_0$ incorporates parameters $\boldsymbol{\lambda}$ , so sensitivity to initial conditions $\boldsymbol{x}_0$ includes PSS parameter sensitivity. A graphical view of such TS idea applied in $\boldsymbol{\Gamma}_1$ can be seen from Fig. 4.

However, the task of obtaining the TS solutions of $\nabla_{\boldsymbol{x}_0}\boldsymbol{x}(t)$ , $\nabla_{\boldsymbol{x}_0}\boldsymbol{y}(t)$ is a complex problem due to TS discontinuity mentioned earlier. Our philosophy here is to firstly consider the smooth HPS between $SE^{(1)}$ and $SE^{(2)}$ that is away from each event, and then to treat jump conditions as the initial conditions for this smooth HPS, and so on. As mentioned earlier, in this paper we assume the HPS at time $t_0$ is already in steady state operation, which indicates that the system TSs at initial time $t_0$ are equivalent to those at prior time $t_{J1}^-$, namely

$$\nabla_{\boldsymbol{x}_0}\boldsymbol{x}(t_{J1}^-)\equiv\nabla_{\boldsymbol{x}_0}\boldsymbol{x}(t_0) \tag{29}$$

$$\nabla_{\boldsymbol{x}_0}\boldsymbol{y}(t_{J1}^-)\equiv\nabla_{\boldsymbol{x}_0}\boldsymbol{y}(t_0) \tag{30}$$

where the initial conditions satisfies

$$\nabla_{\boldsymbol{x}_0}\boldsymbol{x}(t_0)=\boldsymbol{I} \tag{31}$$

$$\boldsymbol{0}=[\nabla'_{\boldsymbol{x}}\boldsymbol{g}^{(0)}+\nabla'_{\boldsymbol{y}}\boldsymbol{g}^{(0)}\nabla_{\boldsymbol{x}_0}\boldsymbol{y}]\,|_{t_0} \tag{32}$$

As indicated in [3], the jump conditions for the sensitivity $\nabla_{\boldsymbol{x}_0}\boldsymbol{x}$ after the event triggering, at time $t_{J1}^+$ are given by

$$\nabla_{\boldsymbol{x}_0}\boldsymbol{x}(t_{J1}^+)=\nabla_{\boldsymbol{x}_0}\boldsymbol{x}(t_{J1}^-)-(\boldsymbol{f}^+-\boldsymbol{f}^-)\nabla_{\boldsymbol{x}_0}t_{J1} \tag{33}$$

where

$$\boldsymbol{f}^+=\boldsymbol{f}(\boldsymbol{x},\boldsymbol{y}^+)\,|_{t_{J1}^+} \tag{34}$$

$$\boldsymbol{f}^-=\boldsymbol{f}(\boldsymbol{x},\boldsymbol{y}^-)\,|_{t_{J1}^-} \tag{35}$$

The sensitivity $\nabla_{\boldsymbol{x}_0}\boldsymbol{y}$ immediately after the event (at time $t_{J1}^+$) are given by

$$\nabla_{\boldsymbol{x}_0}\boldsymbol{y}(t_{J1}^+)=[-(\nabla_{\boldsymbol{y}}\boldsymbol{g}^{(1+)})^{-1}\nabla_{\boldsymbol{x}}\boldsymbol{g}^{(1+)}\nabla_{\boldsymbol{x}_0}\boldsymbol{x}]\,|_{t_{J1}^+} \tag{36}$$

where $\nabla_{\boldsymbol{x}_0}\boldsymbol{x}(t_{J1}^+)$ has been given in (33). Hence, we get the whole information of TS at time $t_{J1}^+$

Similarly, we can use the TSs $\nabla_{\boldsymbol{x}_0}\boldsymbol{x}(t_{J1}^+)$ , $\nabla_{\boldsymbol{x}_0}\boldsymbol{y}(t_{J1}^+)$ to calculate the TSs $\nabla_{\boldsymbol{x}_0}\boldsymbol{x}(t_{J2}^-)$ , $\nabla_{\boldsymbol{x}_0}\boldsymbol{y}(t_{J2}^-)$ at the next event epoch $t_{J2}^-$ , which the TSs are governed by differentiating(15), the second equation of (17) with respect to $\boldsymbol{x}_0$ , we get

$$d\dot{\boldsymbol{x}}\,/\,d\boldsymbol{x}_0=\nabla'_{\boldsymbol{x}}\boldsymbol{f}(t)\nabla_{\boldsymbol{x}_0}\boldsymbol{x}(t)+\nabla'_{\boldsymbol{y}}\boldsymbol{f}(t)\nabla_{\boldsymbol{x}_0}\boldsymbol{y}(t) \tag{37}$$

$$\boldsymbol{0}=\nabla'_{\boldsymbol{x}}\boldsymbol{g}^{(1+)}(t)\nabla_{\boldsymbol{x}_0}\boldsymbol{x}(t)+\nabla'_{\boldsymbol{y}}\boldsymbol{g}^{(1+)}(t)\nabla_{\boldsymbol{x}_0}\boldsymbol{y}(t) \tag{38}$$

where $t\in[t_{J1}^+,t_{J2}^-]$ .

Likewise, by updating the jump condition (33)-(35) , (36)

$$\begin{bmatrix}\nabla_{\boldsymbol{x}_{c0}}\boldsymbol{x}_c & \nabla_{\boldsymbol{z}_0}\boldsymbol{x}_c & \nabla_{\boldsymbol{\lambda}}\boldsymbol{x}_c\\ \nabla_{\boldsymbol{x}_{c0}}\boldsymbol{z}_c & \nabla_{\boldsymbol{z}_0}\boldsymbol{z}_c & \nabla_{\boldsymbol{\lambda}}\boldsymbol{z}_c\\ \nabla_{\boldsymbol{x}_{c0}}\boldsymbol{\lambda}_c & \nabla_{\boldsymbol{z}_0}\boldsymbol{\lambda}_c & \nabla_{\boldsymbol{\lambda}}\boldsymbol{\lambda}_c\end{bmatrix}_{t_{J1}^+}=\begin{bmatrix}1&0&0\\0&1&0\\0&0&1\end{bmatrix}+\boldsymbol{f}^*\begin{bmatrix}\nabla_{\boldsymbol{x}_{c0}}t_{J1} & \nabla_{\boldsymbol{z}_0}t_{J1} & \nabla_{\boldsymbol{\lambda}}t_{J1}\\0&0&0\\0&0&0\end{bmatrix} \tag{43}$$

$$\begin{bmatrix}\nabla_{\boldsymbol{x}_{c0}}\boldsymbol{y}\\ \nabla_{\boldsymbol{z}_0}\boldsymbol{y}\\ \nabla_{\boldsymbol{\lambda}}\boldsymbol{y}\end{bmatrix}'_{t_{J1}^+}=\left[\nabla_{\boldsymbol{y}}\boldsymbol{g}^{(1+)}\right]^{-1}_{t_{J1}^-}\begin{bmatrix}\nabla_{\boldsymbol{x}_{c0}}\boldsymbol{g}^{(1+)}\\ \nabla_{\boldsymbol{z}_0}\boldsymbol{g}^{(1+)}\\ \nabla_{\boldsymbol{\lambda}}\boldsymbol{g}^{(1+)}\end{bmatrix}'_{t_{J1}^+}\begin{bmatrix}\boldsymbol{f}^*\nabla_{\boldsymbol{x}_{c0}}t_{J1}-1 & \boldsymbol{f}^*\nabla_{\boldsymbol{z}_0}t_{J1} & \boldsymbol{f}^*\nabla_{\boldsymbol{\lambda}}t_{J1}\\0&1&0\\0&0&1\end{bmatrix} \tag{44}$$

$$\nabla_{\boldsymbol{\lambda}}\boldsymbol{y}(t_{J1}^+)=\{-\left[\nabla_{\boldsymbol{y}}\boldsymbol{g}^{(1+)}\right]^{-1}\left[\nabla_{\boldsymbol{\lambda}}\boldsymbol{g}^{(1+)}-\nabla_{\boldsymbol{\lambda}}\boldsymbol{g}^{(1+)}\boldsymbol{f}^*\nabla_{\boldsymbol{\lambda}}t_{J1}\right]\}_{t_{J1}^+} \tag{46}$$

as

$$\nabla_{x_0} x(t_{J2}^{+}) = \nabla_{x_0} x(t_{J2}^{-}) - (f^{+} - f^{-})\nabla_{x_0} t_{J2} \tag{39}$$

where

$$f^{+} = f(x, y^{+})\,|_{t_{J2}^{+}} \tag{40}$$

$$f^{-} = f(x, y^{-})\,|_{t_{J2}^{-}} \tag{41}$$

$$\nabla_{x_0} y(t_{J2}^{+}) = [-(\nabla_y g^{(2+)})^{-1}\nabla_x g^{(2+)}\nabla_{x_0} x]\,|_{t_{J2}^{+}} \tag{42}$$

then, we can obtain the whole information of TS (i.e., $\nabla_{x_0} x(t_{J2}^{+})$ , $\nabla_{x_0} y(t_{J2}^{+})$) at time $t_{J2}^{+}$ .

By applying the above compact equations into obtaining the TSs for PSS parameter case, we can get (43), (44) (after arrangement) , where $f^{*} = f^{-} - f^{+}$ . Thus, we have

$$\nabla_{x_{c0}} \lambda_c(t_{J1}^{+}) = f^{*}\nabla_{\lambda} t_{J1} \tag{45}$$

Therefore, we can use(37), (38), (45), (46) to calculate $\nabla_{x_0} x(t_{J2}^{+})$ , $\nabla_{x_0} y(t_{J2}^{+})$ . After this, we can get the TSs $\nabla_{x_0} x(t)$ , $\nabla_{x_0} y(t)$ , where $t \in [t_{J2}^{+}, t_f]$ by

$$d\dot{x} \,/\, dx_0 = \nabla_x' f(t)\nabla_{x_0} x(t) + \nabla_y' f(t)\nabla_{x_0} y(t) \tag{47}$$

$$0 = \nabla_x' g^{(2+)}(t)\nabla_{x_0} x(t) + \nabla_y' g^{(2+)}(t)\nabla_{x_0} y(t) \tag{48}$$

### C. Numerical Calculations of TSs and CGM

Up to now, we have collect all the TS information which can be used to calculate gradient information $\nabla_{\lambda} J(x, t_f)$ in (26). In fact, for large HPS the number of equation is huge, and the computational cost may be high. Fortunately, as indicated in [3], by using an implicit numerical integration technique such as trapezoidal integration, the computational burden of obtaining the TSs can be reduced a lot. Here we are not going to focus in this numerical integration, whose main idea is to approximate differential equations using proper difference equation. And then form a set of nonlinear algebraic equations, which can be easily solved by Newton's Method. It is worth mentioning that both the state variable $\omega_i$ and TS information $\nabla_{\lambda}\omega_i$ can be directly obtained as by-products of the numerical simulations/calculations. We will use them in the following optimization of $J(x, t_f)$ by applying CGM.

As illustrated in [22], the basic idea of gradient-based optimization methods is to find a descent direction $d_k$ and corresponding positive step $\alpha_k$ with an initial value $\lambda_0$ , then it iterates itself up to $n$ times, which satisfies

$$\lambda_{k+1} = \lambda_k + \alpha_k d_k, \quad \alpha_k > 0 \tag{49}$$

$$\nabla_{\lambda}' J(\lambda_k) d_k < 0 \tag{50}$$

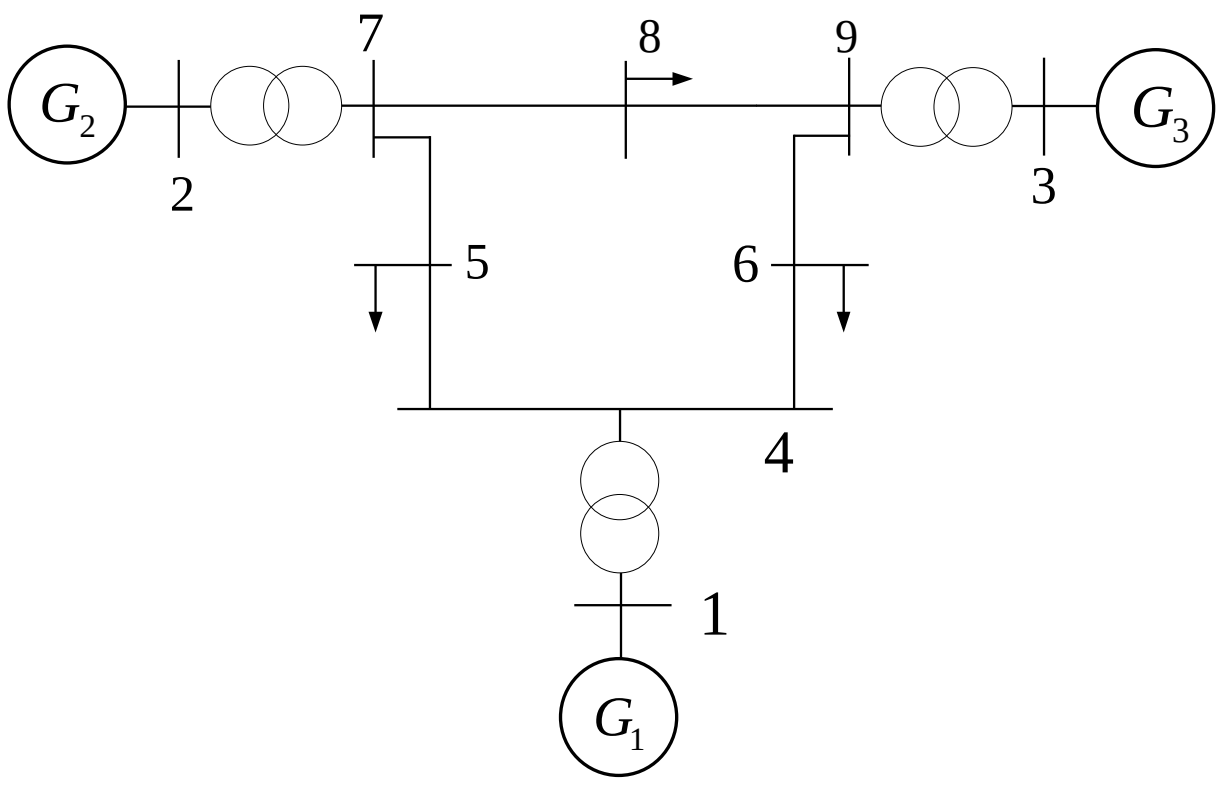

Fig. 5. Diagram of IEEE three-machine-nine-bus standard test system.

TABLE I
COMPARISON BETWEEN PSASP AND TS OPTIMAL METHODS FOR PSS PARAMETERS TUNING OF IEEE THREE-MACHINE-NINE-BUS STANDARD TEST SYSTEM

| States | $K_s$ | $T_1$ | $T_2$ | $T_3$ | $T_4$ |
|---|---|---|---|---|---|
| PSAP optimal | 7.5 | 0.174 | 0.050 | 0.174 | 0.050 |
| TS optimal ($G_2$) | 7.52 | 0.672 | 0.041 | 0.672 | 0.041 |
| TS optimal ($G_3$) | 7.51 | 0.460 | 0.084 | 0.460 | 0.084 |

$$J(\lambda_k + \alpha_k d_k) < J(\lambda_k), \ \forall k = 1, ..., n-1 \tag{51}$$

where the iteration terminates by certain feasible criteria, such as $\left\|\nabla_{\lambda} J(\lambda_k)\right\| < \varepsilon$ ( $\varepsilon$ is a very small positive real number).

Here, the steepest descent method could be used to select $d_k$ , which equals to $-\nabla_{\lambda} J(\lambda_k)$ . However, this gradient method may meet be quite slow when $\lambda_k$ approach to its optimal value. The CGM could be regarded as a modified steepest descent method, which by updating its descent direction $d_k$ as a linear combination of previous descent direction $d_{k-1}$ and its latest gradient information $\nabla_{\lambda} J(\lambda_k)$ as follows.

$$d_{k+1} = -\nabla_{\lambda} J(\lambda_{k+1}) + \beta_{k+1} d_k \tag{52}$$

where $\beta_{k+1}$ can be obtain by using Powell-Fletcher-Reeves Rule, namely

$$\beta_k = \frac{\nabla_{\lambda}' J(\lambda_{k+1})\left[\nabla_{\lambda} J(\lambda_{k+1}) - \nabla_{\lambda} J(\lambda_k)\right]}{\nabla_{\lambda}' J(\lambda_{k+1})\nabla_{\lambda} J(\lambda_k)}, \quad \forall k = 1, ..., n-1 \tag{53}$$

And for stepsize $\alpha_k$ , we use Armijo Rule as

$$J(\lambda_k) - J(\lambda_k + \rho_{m_k} s d_k) \geq -\sigma\beta_k s \nabla_{\lambda}' J(\lambda_k) d_k \tag{54}$$

where $\rho \in [0,1]$ , $\sigma \in [0,1]$ and $m_k$ is the first nonnegative

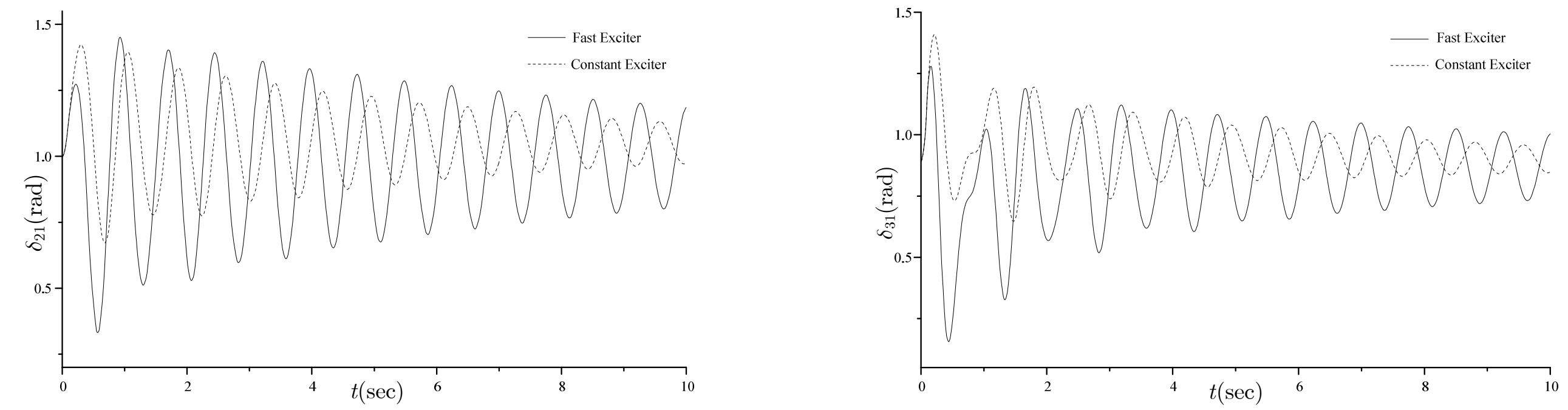

(a) (b)

Fig. 6. Power angle curves under fast and constant exciters. (a) $\delta_{21}$ ; (b) $\delta_{31}$ .

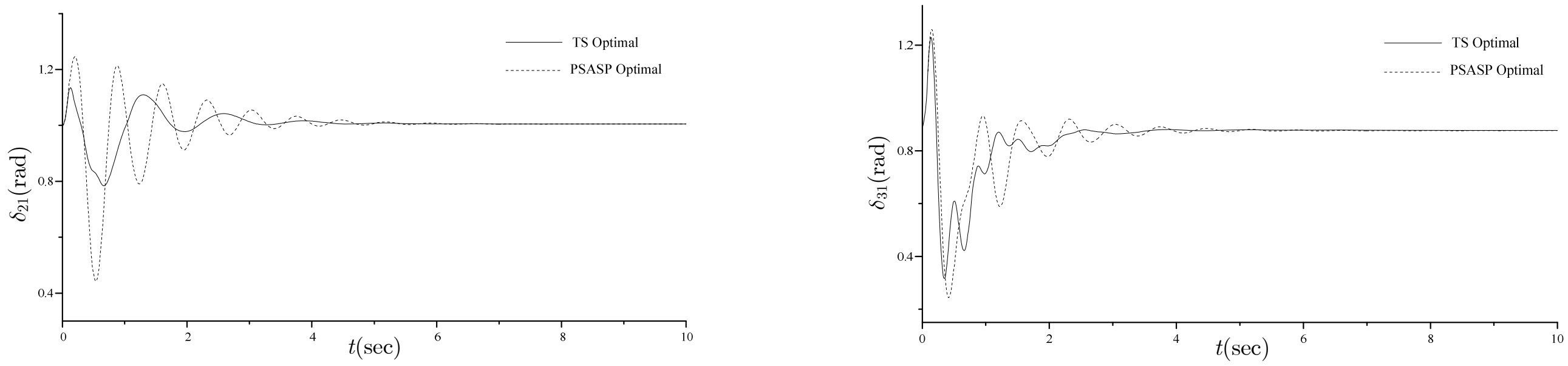

(a) (b)

Fig. 7. Power angle curves under PSASP and TS optimal methods. (a) $\delta_{21}$ ; (b) $\delta_{31}$ .

integer that satisfies the above inequality.

Therefore, by applying CGM to the optimization problem(25), we can eventually get the optimal parametric tuning rule for each PSS associated with generators in the HPS.

## IV. Application to IEEE Standard Test System

In this Section, we will apply the optimal parametric tuning method of PSS developed in this paper. The test system is the famous Anderson three-machine-nine-bus standard test system as depicted in Fig. 5, which is also an IEEE standard test one [23]. PSSs are installed for $G_2$ and $G_3$ while $G_1$ operates without any PSS. The system frequency selected as 60 Hz, the simulation time is set as 10 second. The trigger events applied here belongs to switching events, where the 9-th Bus is in three-phase to ground short-circuit fault that occurs at time 0 and subside at 0.1 second.

### *A. Transient Process without PSS*

As mentioned earlier, the lack of damping in the power system is one of the main reasons for low-frequency oscillation. However, the fast exciter which is designed to improve the response of generator could deteriorate such damping that lead to sever oscillation after large disturbance. First, we would like to validate this case by neglecting PSS for $G_2$, $G_3$ and in Fig. 5. The exciter parameter are chosen as $K_a$=24, $T_a$=0.05. Here, we use power angles $\delta_{21}$, $\delta_{31}$ with respect to $G_1$, which can be used to characterize operational behavior of $G_2$, $G_3$. It should be emphasized that $\delta_{21}$, $\delta_{31}$ also belong to dynamic state variable $\boldsymbol{x}$, and its first derivative equals to $\omega_i$ .

Fig. 6 show that there is significant amplitude increasing of the power angle oscillation after using fast exciter for both $G_2$ and $G_3$ . To solve this problem, we will apply the optimization-based PSS parametric tuning method to inhibit oscillations under large disturbance.

### *B. Transient Process with optimal PSS*

For the IEEE standard test system shown in Fig. 5, we install fast exciter with PSS for $G_2$ and $G_3$, where $T_w$=10, $K_a$=24, $T_a$=0.05, and optimize other parameters $K_s$, $T_1$, $T_2$ using TS approach. $G_1$ is still using constant exciter. Here, we will compare our obtained PSS optimal parameters with those determined by Power System Analysis Software Package (PSASP, based on small-signal analysis) as shown in Table I, and plot the corresponding power angles $\delta_{21}$, $\delta_{31}$ as depicted in Fig. 7 (a), (b). Obviously, the proposed TS optimal PSS parametric tuning method has a better effect for damping oscillations than PSASP optimal method, when the power system undergoes a large disturbance.

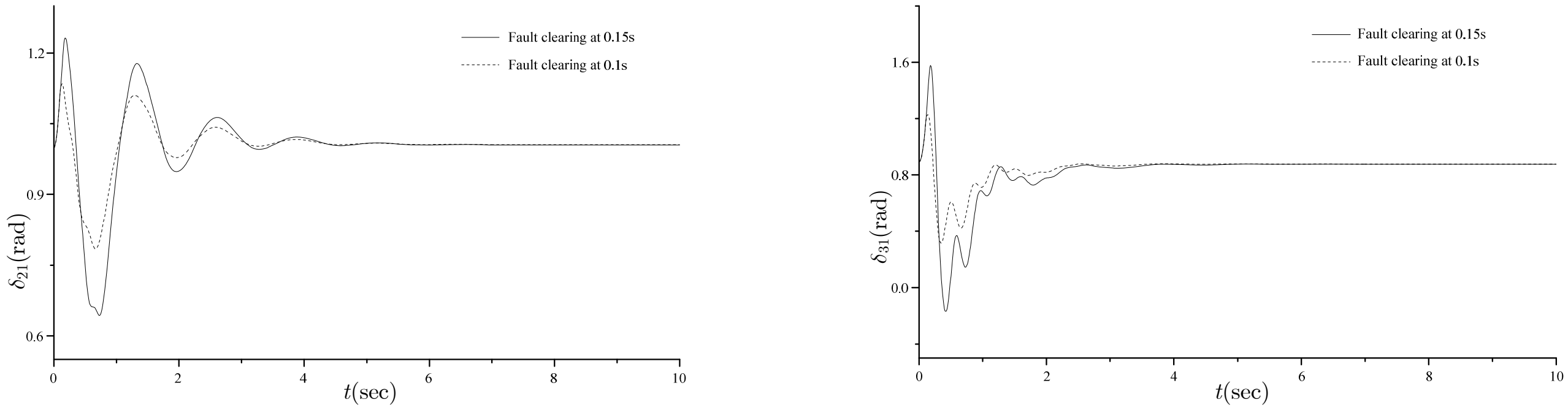

(a) (b)

Fig. 8. Power angle curves with different fault clearing at 0.15s and 0.1s. (a) $\delta_{21}$ ; (b) $\delta_{31}$ .

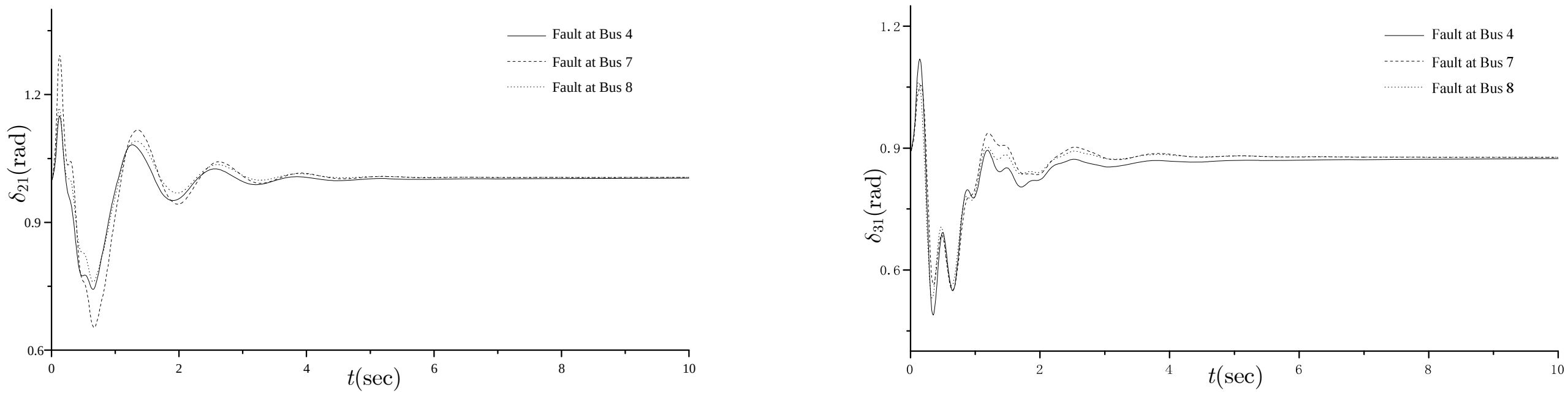

(a) (b)

Fig. 9. Power angle curves with fault under different buses.. (a) $\delta_{21}$ ; (b) $\delta_{31}$ .

Moreover, to illustrate our previous discussions on discrete-events that occur at different epochs with different influence on the change of power networks, we apply the optimal PSS tuning parameters to different jumping time and different switching events, respectively.

First, let the three-phase to ground short-circuit fault that occurs at time 0 and subside at 0.15 second. The power angles $\delta_{21}$, $\delta_{31}$ are shown in Fig. 8 (a), (b), which indicate that the previous optimal PSS parameters doesn't work very well for a prolong switching event. The reason is that when the interval between two switching events become larger, the TSs $\nabla_{\boldsymbol{x}_0}\boldsymbol{x}(t_{\bar{J}2})$ , $\nabla_{\boldsymbol{x}_0}\boldsymbol{y}(t_{\bar{J}2})$ at the next event epoch $t_{\bar{J}2}$ are different from the previous case with a shorter event epoch $t_{\bar{J}2}$ . Thus, the gradient information of $\nabla_{\boldsymbol{\lambda}}J(\boldsymbol{x},t_f)$ obtained from the above Tss should also be different, which lead to the previous optimal parameters become suboptimal now.

Moreover, let's keep the fault clearing at 0.1 second and change the bus number which has three-phase to ground short-circuit fault to 4, 7, 8, respectively. The corresponding power angles are shown in Fig. 9 (a), (b), which indicate that the previous optimal PSS parameters doesn't work very well for different switching events. This leads to the original optimal PSS parameters obtained in one switching event become suboptimal at another. The reason is reliant on the changes of power network algebraic equations are different under different switching events. Thus according to the detailed explanation in Section III, the gradient information of $\nabla_{\boldsymbol{\lambda}}J(\boldsymbol{x},t_f)$ obtained from Tss will also be different so that the optimal parameter become suboptimal.

From the above examples, we can see that the damping effects are almost all acceptable in terms of inhibit large disturbance. However, when the actual power system becomes larger, it is very difficult for all types of faults to be preferably suppressed by adjust PSSs. One solution is to implement an online optimization for PSS parameters, namely to automatically calculate the most optimal parameters so that the power system can recover quickly. Nevertheless, this requires rather high computational speed, which is absent for TS-based PSS parameter optimization method. In addition, the online optimization could also involve the site selection for PSS installation, and further research may be required to achieve these requirements, which is not the focus of the interest herein.

## V. Conclusion

In this paper, the optimal parametric tuning method of PSS in HPS is studied from the viewpoint of transient trajectory sensitivities both theoretically and numerically. Concrete definitions of switching and reset events are provided so as to formulate triggering hypersurfaces to derive the model of HPS, which are capable of revealing the transient behavior of trajectory sensitivities. It is shown that the discontinuity and discrete states are major obstacles to analyze the constraints of this optimization problem. The gradient information of the objective function with respect to decision variables can be obtained by means of calculating trajectory sensitivities of state variables to the PSS parameters, where a conjugate gradient method has been implemented to optimize these PSS parameters. In this paper, the core of computation is to solve the state equations as well as its adjoint TS equations. The application results for IEEE standard test systems show that the proposed optimal parametric tuning approach can effectively damp the system oscillation caused by large disturbance.

## ACKNOWLEDGMENT

The author would like to thank Miss. Yuan Zhang (Xi'an Jiaotong University, China), Prof. Yannis Paschalidis (Boston University), for their valuable discussions and insightful feedback. I also wish to thank Grace Dai for her kind supports at different stages of preparing this paper.